# Comparative Analysis of User Behavior of Dock-Based vs. Dockless Bikeshare and Scootershare in Washington, D.C.

**Kiana Roshan Zamir** (Corresponding author)
Ph.D. Candidate, Department of Civil and Environmental Engineering
University of Maryland, College Park, MD 20742, USA
Email: kianarz@umd.edu

**Iryna Bondarenko**
Master of Community Planning, School of Architecture, Planning, and Preservation
University of Maryland, College Park, MD 20742, USA
Email: isbondarenko@gmail.com

**Arefeh Nasri**
Faculty Research Scientist
National Center for Smart Growth Research and Education, Maryland Transportation Institute,
University of Maryland, College Park, MD 20742, USA
Email: aanasri@umd.edu

**Stefanie Brodie**
Research Program Specialist
District Department of Transportation, 55 M St SE, Washington, DC 20003, USA
Email: stefanie.brodie@dc.gov

**Kimberly Lucas**
Supervisory Transportation Management Planner
Bicycle and Pedestrian Program Specialist
District Department of Transportation, 55 M St SE, Washington, DC 20003, USA
Email: kimberly.lucas@dc.gov

**ABSTRACT**

In 2017, dockless bikeshare systems were introduced in the United States, followed by dockless scootershare in early 2018. These new mobility options are expected to complement the existing station-based bikeshare systems, which are bound to static origin and destination points at docking stations. The three systems attract different users with different travel behavior mobility patterns. The present research provides a comparative analysis of users' behavior for these three shared mobility systems during March-May 2018 in the District of Columbia. Our study identifies similarities/differences between the two systems aiming for better planning, operating, and decision-making of these emerging personal shared mobility systems in the future. It uses logistic regression and random forest modeling to delineate between "member" behavior, which aligns most closely with commuter behavior, and "casual" behavior that represents more recreational behavior. The results show that 63.8% of dockless bike users and 69.6% of dockless scooter users demonstrated "member" behavior, which is slightly lower than the actual percentage of trips made by members within the conventional bikeshare system (73.3%). Dockless systems users also showed to have short trip durations similar to conventional bikeshare system's registered members, with no significant difference between trips during weekdays and weekends. Overall, this study provides a methodology to understand users' behavior for the dockless bikeshare system and provides sufficient evidence that these new shared mobility systems can potentially make positive contributions to urban multi-modal infrastructure by promoting bicycle usage for urban daily travel.

## 1. Introduction

Shared mobility services continue to grow and evolve due to technological advancements around the world. Among which, bikeshare program is one that has been expanding around the world within the last several years due to its various social, environmental, and health benefits (*DeMaio, 2009; Shaheen et al., 2010*). In the United States alone, bikeshare programs are supported by different operators working under city municipalities and transportation agencies and now offer services in more than 70 metropolitan areas, cities, and small towns. The success of these programs was observed through the millions of trips taken daily nationwide using the systems. However, there are some limitations associated with these new shared mobility services as well, such as fixed stations, limited capacity in each station, and the concentration of stations mostly in city centers and high-density areas, which all limit the network coverage and users' accessibility to the stations. Many of the previous studies have confirmed the influence of accessibility of bikes in the station-based bikeshare programs on the usage and membership of these systems *(Fuller et al., 2011; Molina-Garcia et al., 2013; and Fishman et al., 2015),* The new generation of bikeshare emerged as dockless systems to cope with the aforementioned limitations. While traditional dock-based bikeshare systems require trips to begin and terminate at static docking stations, the new generation employs a dockless model, where bikes can be unlocked, used, and returned anywhere in the city depending on the user's origin/destination locations. In July 2017, the first dockless bikeshare pilot program was launched in the United States in the city of Seattle. Dockless vehicles, which include both bikes and scooters, now exist alongside traditional dock-based systems in many large municipalities across the United States. It is estimated that around1.4 million trips were made using dockless bikes in the United States in 2017 *(NACTO, 2018)*. The conventional and dockless systems are expected to complement each other toward promoting and facilitating bikeshare usage. However, in order to assess the success of these systems, one important factor is to understand the behavior of users for each of these systems for better planning, operation, and management of the systems in the future. This study attempts to analyze the similarities/differences of the users' behavior between staion-based and dockless systems to help inform municipal planning and policy decisions around bikes and other emerging personal shared mobility systems.

The contributions of this study to the existing literature are twofold. First, we compared three differing modes of shared mobility, namely dockless electric scooter, dockless bike, and dock-based bikeshare systems within Washington, D.C as a case study area. This study is one of the firsts to address the similarities and differences of travel pattern among users of these systems, such as their start/end location distribution, temporal trip distribution, and trip duration. Findings of this paper can help operators of dock-based and dockless vehicle sharing systems to improve their operations by providing insight into dockless data. Also, the results of this research may help cities who already have or plan to launch bikeshare programs. The following section provides a review of previous literature on both conventional and dockless bikeshare systems and identifies research gaps in this area. Next, data collection and processing steps are presented along with brief descriptive statistics of the two systems in the District of Columbia, followed by a comprehensive comparison of the users' behavior of dockless systems with "casual users" and "members" of dock-based systems. Random forest and logistic regression are used to train a model using Capital Bikeshare data that can classify the trips into trips taken by casual users and members. "Casual" riders represent users that purchased single trips or daily or 3-day passes, and "members" have purchased annual bikeshare membership. By applying these models to the dockless data, this study will investigate whether behavior pattern of dockless vehicles' users is more similar to Capital Bikeshare casual users or members. The study concludes with the interpretation and discussion of the results and recommendations to the systems' operators and decision-makers.

2. Literature review

Numerous studies investigated different aspects of conventional bikeshare systems such as impacts of bikeshare *(DeMaio, 2009; Shaheen et al., 2010; Woodcock et al., 2014; El-Geneidy, 2016)*, users' characteristics *(Fishman, 2016; Buck et al., 2013)*, demand *(Nasri, et al., 2018; Younes et al., 2019)*, interactions with other modes *(Bachand-Marleau, 2011; Martin and Shaheen, 2014; Hamilton and Wichman, 2018; Barber and Starrett, 2018; Ma et al., 2018)*. However, research on the new dockless bikeshare and scootershare systems is scarce. There are studies on dockless bicycles as a component of a multi-modal transportation system *(Zhou and Zhang, 2018),* as well as research and recommendations for better rebalancing practices *(Liu and Xu, 2018; Pal and Zhang, 2017; Pan et al., 2018)*, demand prediction *(Ai et al., 2018; Liu et al., 2018)*, their spatial distributions *(McKenzie, 2018),* and equity analysis *( Mooney et al., 2019).* While it is important to investigate the similarities and differences between the two systems for future planning and policy purposes, there are, to the best of authors' knowledge, no analyses done in the past about the detailed travel pattern of users of major dockless systems, and travel behavior comparison of dockless vs. station-based systems users, especially within the United States. Moreover, this paper which is an extended version of Zamir et al. (2019) is the first paper analyzing the travel pattern of dockless scootershare programs in the United States.

Several studies analyzed the dockless systems' user behavior in China. For instance, a study by Du and Cheng (2018) identified factors influencing the travel pattern of dockless bike sharing users in Nanjing, China using a survey and by fitting a multinomial logit model. Travel pattern of users were divided into three categories: origin to destination pattern (ODP), travel cycle pattern (TCP), and transfer pattern (TP). The results showed employees and students mostly choose TCP and ODP, especially for shorter trips. They also found that price and the existence of malfunctioning bikes have high impacts on the usage of dockless systems. Ai et al. (2018) also compared the traveler's transfer tolerance of walking to users of dockless bikeshare in Chengdu, China, and suggested that pedestrians are more tolerant to the environment than dockless users. Other studies from China showed that dockless bicycles were used at the same time of day as the existing dock-based system and had nearly the same trip durations *(Yang et al., 2017; Li et al. 2018).* They found that users of the dockless bikeshare systems made short-distance trips during rush hours, mostly for commuting or studying, the same as other commuters. The launch of a dockless system, in addition to an existing dock-based one, allowed users to buy single trips without registration, while the dock-based system did not provide this option *(Li et al., 2018).* Chen at al. (2018) analyzed the users' behavior and factors influencing systems' usage frequency for both conventional and dockless systems in Hangzhou, China, and found that both of the systems are dominantly used by males and users younger than 35. Most of them don't have cars or e-bikes and their top 3 trip purposes are commuting, going to school, and for leisure. The result of ordinal logistic regression showed that for dockless users having high cellphone data and for station-based systems the education background of master/Ph.D has a substantial role in their frequency of usage. The result also indicated dockless trip purpose is very flexible whereas station-based systems tend to be used for high-frequency trips (i.e., commuting.)

In Europe, a study on dockless Bikeshare system showed 50 percent of users have used bikes less than 5 times a year and 80 percent of total trips are made by only 20 percent of users. The temporal usage of dockless bike showed three clear peaks; 8 a.m.-9 a.m., 12-2 p.m., and 6-8 p.m. for the weekday which is somewhat different from the pattern observed in the Washington, DC. They also suggested that precipitation has a significant impact on the usage of dockless systems that lasts even after the rain *(Reiss and Bogenberger, 2015)*.

In the United States, a study by Pal at al. (2018) analyzed the dockless trip data from Tampa campus of the University of Florida and found that most of the trips took place during the fall season. Also, the number of trips during weekdays was substantially higher than the weekends and the usage was at its highest around 1 p.m. on weekdays. Virginia Tech researchers conducted a study on dockless bikeshare in Washington, D.C. in September 2017- January 2018 and showed that peak hours of dockless bikeshare usage were different from typical commuting hours. The morning peak for dockless bikeshare users started at 9 a.m. and dropped at 11 a.m., one hour later than that of CaBi users. The afternoon peak was longer in duration, starting at 12 p.m. and ending at nearly 8 p.m. The chosen connections were also different for CaBi compared to dockless system's users. Dockless bicycles' riders rode slightly different paths than CaBi users, and relatively, dockless trips were more distributed outside the Central Business District (CBD) than dock-based trips (VirginiaTech, 2018). Based on GPS data from 94 CaBi bicycles, Wergin and Buehler (2017) found that casual riders started their trips from different locations than members, and followed different routes. Members showed strong commuting patterns, starting their trips in residential or mixed-use neighborhoods, and the casual riders' trips were mostly concentrated around the National Mall. Also, while members often kept their trips short and direct, casual riders made longer trips (on average three times longer in duration than members) and made frequent detours. As a result, the distance of an average casual rider's trip was twice as long as the distance of an average member's trip *(Wergin and Buehler, 2017)*. The fact that registered bikeshare users made shorter trips, was revealed also by Khatri et al. *(2016)*. Another study done by Noland et al. (2016) analyzed the New York dock-based bikeshare users' behavior and found that unlike casual users, members tend to start their trips in high residential and employment density areas and mixed-use neighborhoods. However, neither of the casual nor members' trips were not located in solely residential land use neighborhoods.

In 2018, Virginia Tech researchers did an intercept survey of dockless bikeshare users. The survey was conducted in winter and involved only 49 riders. With these limitations acknowledged, the study found that dockless bikeshare users were more racially diverse, and had more female riders compared to CaBi members. The results of this survey also indicated that only 34% of dockless responders made more than five trips in the past 30 days whereas this number was 58% for CaBi members *(VirginiaTech, 2018)*.

Using data from dockless bikeshare and scootershare systems in Washington, D.C., the present study tried to fill in the aforementioned gaps in the literature on the users' characteristics and behavior of dockless bikes and scooters by investigating the travel patterns of the systems' users and providing a better understanding of these new emerging modes through a comparative analysis of the two conventional and dockless shared mobility systems.

## 3. Data collection and processing

### 3.1. Capital Bikeshare

District of Columbia's SmartBike system was launched in 2008 as the first bikeshare program in the United States and fast became very popular among both residents and tourists. At the beginning, it consisted of 120 bicycles distributed in ten stations across the district. In 2010, the Capital Bikeshare (CaBi) system replaced the old program and expanded bikeshare system to cover the entire Washington, D.C. metropolitan area. Capital Bikeshare has since grown to 507 stations and 5,000 bicycles in the District of Columbia, Virginia (Alexandria, Arlington, Fairfax), and Maryland (Montgomery County and Prince George's County) *(Capital Bikeshare website, 2018)*. The number of trips has also increased tenfold, reaching 350,000 trips per month and more than three million trips per year in 2017. Around 80% of the CaBi trips are made by registered members, who purchased the annual membership for $85 granting unlimited trips on the shared bicycles for up to 30 minutes. While they receive free ride for the first 30 minutes, the system will charge them for $1.5 for the second 30 minutes of the ride, and the fee increases for each subsequent 30-minute period. The CaBi's trip data was accessed and downloaded from their open system data website from March 2018 through May 2018. Descriptive statistics of the data shows that the majority of the annual members are males (61%), white (54%), and between 20-39 years of age (67%) *(Bikeshare Monthly Report, 2016)*. Nearly 20% of trips are made by casual users who purchased a three-day membership ($17), single-day membership ($7 for key holders or $8 for those who buy a pass at the station), or single trips ($2 per ride). An earlier survey on CaBi's casual members, done by Virginia Tech in 2012, showed that the typical casual users were white females between 25 and 34 years of age, frequent cyclists, and domestic or international tourists *(Virginia Tech, 2012)*. At the same time, CaBi's monthly performance reports for 2017 and 2018, prepared by *Alta Planning* and *Motivate*, show that 50-80% of casual trips are made by Washington, D.C.'s residents *(Capital Bikeshare Monthly Reports, 2010-2016)*.

### 3.2. Dockless operators

In September 2017, the District Department of Transportation launched a dockless bikeshare pilot program. During the fall season, they issued permits to five dockless bike operating companies (i.e., JUMP, LimeBike, Mobike, ofo, and Spin). Unlike Capital Bikeshare, these bikes do not need to be returned at docking stations and can be left anywhere in the District's public right-of-way (excluding National Park Service and Federal lands) as long as they do not obstruct roadways and pedestrian walkways. Each of these companies was allowed to operate up to 400 vehicles in the District during the demonstration period. In March 2018, two more companies entered the pilot and introduced electric scootershare to the District (namely Skip and Bird), and one of the existing dockless bikeshare operators (i.e., LimeBike) added electric scooters to its fleet as well. In spring 2018, most dockless bicycles' and scooters' operators sold single rides ($1-2 per 30 minutes or per 1 hour). Spin offered monthly and annual membership, and LimeBike started offering monthly memberships for bicycles only. The data from six dockless companies operating in the District (LimeBike, Mobike, ofo, Spin, Skip, and Bird) was retrieved from the companies for the same period (March-May 2018). This study did not use any application programming interface (API) data and data used is directly from the dockless companies.

3.3. Data processing

To avoid inclusion of incorrect records in the analysis, we have done data cleaning in several steps for both CaBi and the dockless operators' data to make sure the data includes correct trip records for the analysis. During our data cleaning process, the following trips were excluded:

- Trips that were marked as canceled;
- Trips that lasted less than 60 seconds;
- Trips that lasted less than 120 seconds and had the same start and end location;
- Trips that lasted longer than 24 hours; and
- Trips that started or ended outside of the District of Columbia boundaries.

The final dataset for the conventional bikeshare system contained 816,980 trips from March 2018 to the end of May 2018 (599,473 trips taken by registered members and 217,507 trips taken by casual riders). The dockless system data contains 71,590 trips made by dockless bikes and 187,909 trips made by scooters for the same time period of March-May 2018. Data include start/end time, start/end geocoded location information (station for conventional bikeshare trips and exact location for dockless bikeshare trips), and vehicle number and type of membership for each trip (for conventional bikes only).

4. Descriptive analysis results

4.1. Day of Week

Distribution of trips during the week shows different patterns for observed shared vehicles. Figure 1 shows that Capital Bikeshare registered members did most of their trips during weekdays with a peak on Thursdays and clear decline on Saturdays and Sundays. In contrast, most of the trips by casual riders are made during weekends. Dockless bikeshare users did not show sharp decline or increase on weekends. Most of the dockless bikeshare trips were made in the second half of the week with a peak on Saturdays, similar to CaBi casual riders. Scooter trips had a peak on Thursdays (similar to CaBi registered members), while their percentage of trips did not drop on weekends as much as CaBi registers members.

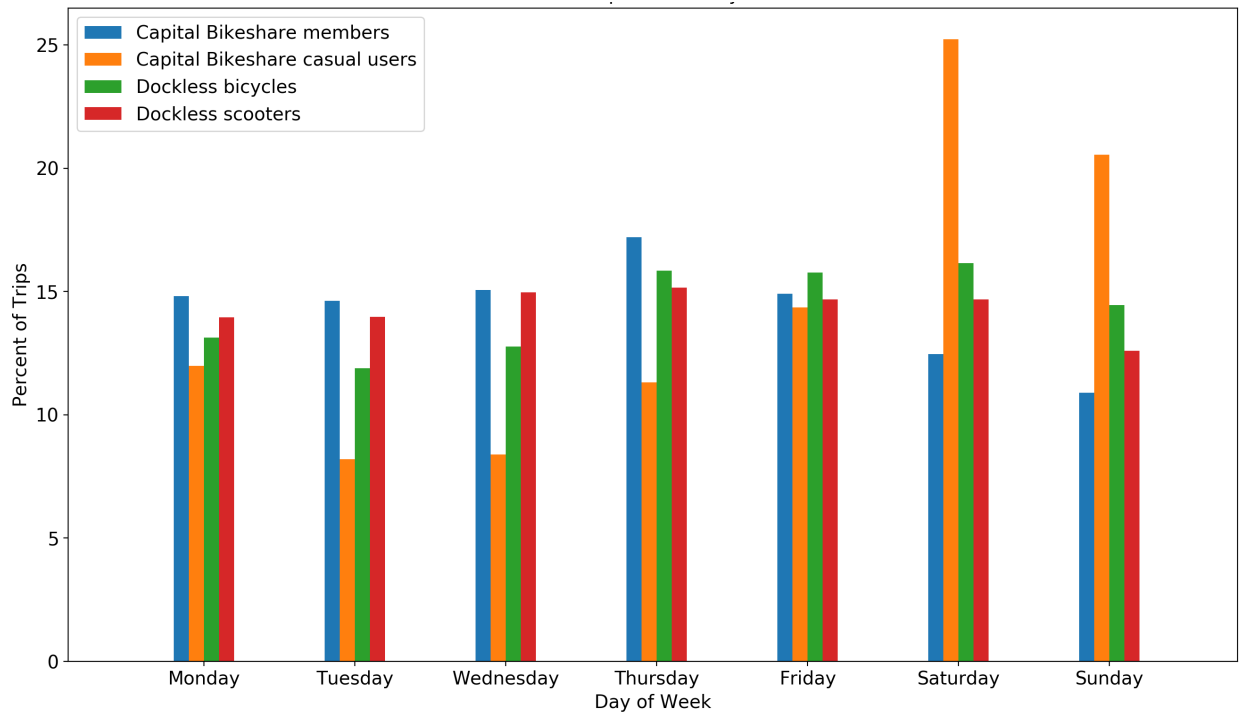

**FIGURE 1 Distribution of trips for each mode by day of week**

## 4.2. Time of day

Peak periods for the types of riders compared in this study differed during the working days and were almost the same on weekends. Figure 2 presents temporal distributions of trips on weekdays vs. weekends for both CaBi and dockless system riders and separated by user type. Four curves are generated on each graph; CaBi registered users, CaBi casual riders, dockless bike riders, and dockless scooter riders. As indicated in Figure 2-a, CaBi members had clear peak hours at 8-9 a.m. and at 5-6 p.m. with a smaller afternoon peak at 12-1 p.m. Nearly 30% of all member trips took place during these three hours. Casual CaBi riders, however, did not have any morning peak. Starting at 6 a.m., the number of casual rides gradually increased reaching the highest point at 5 p.m., and then precipitously dropped.

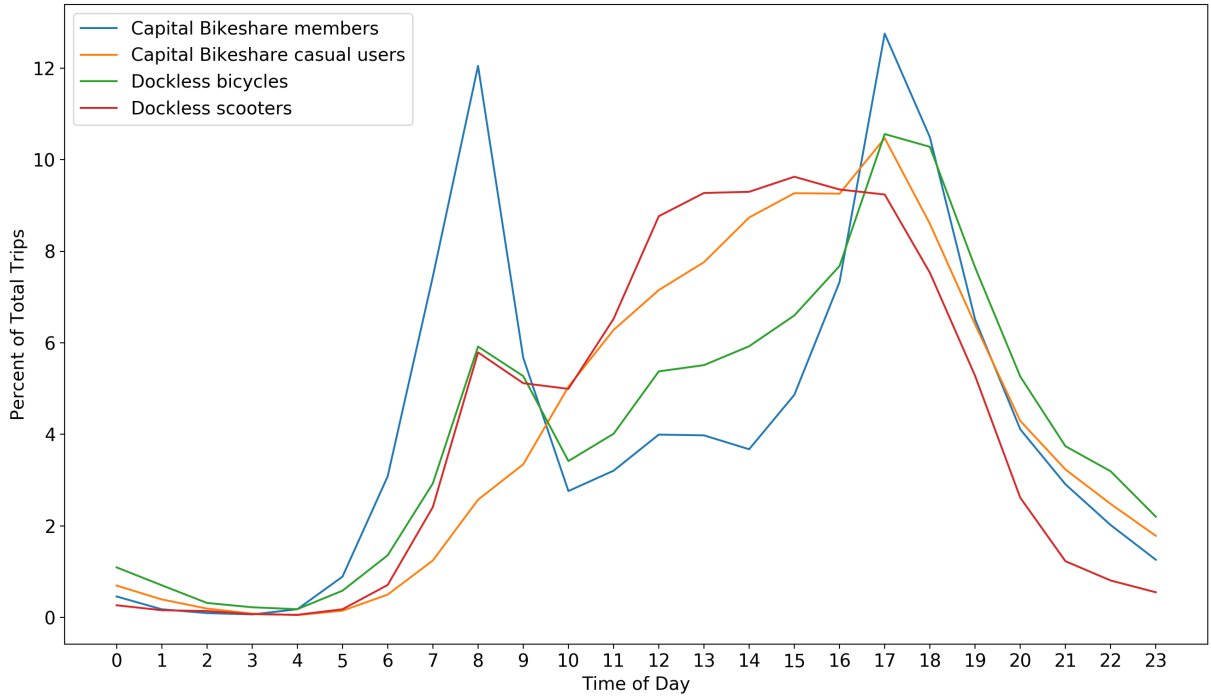

a) Weekdays

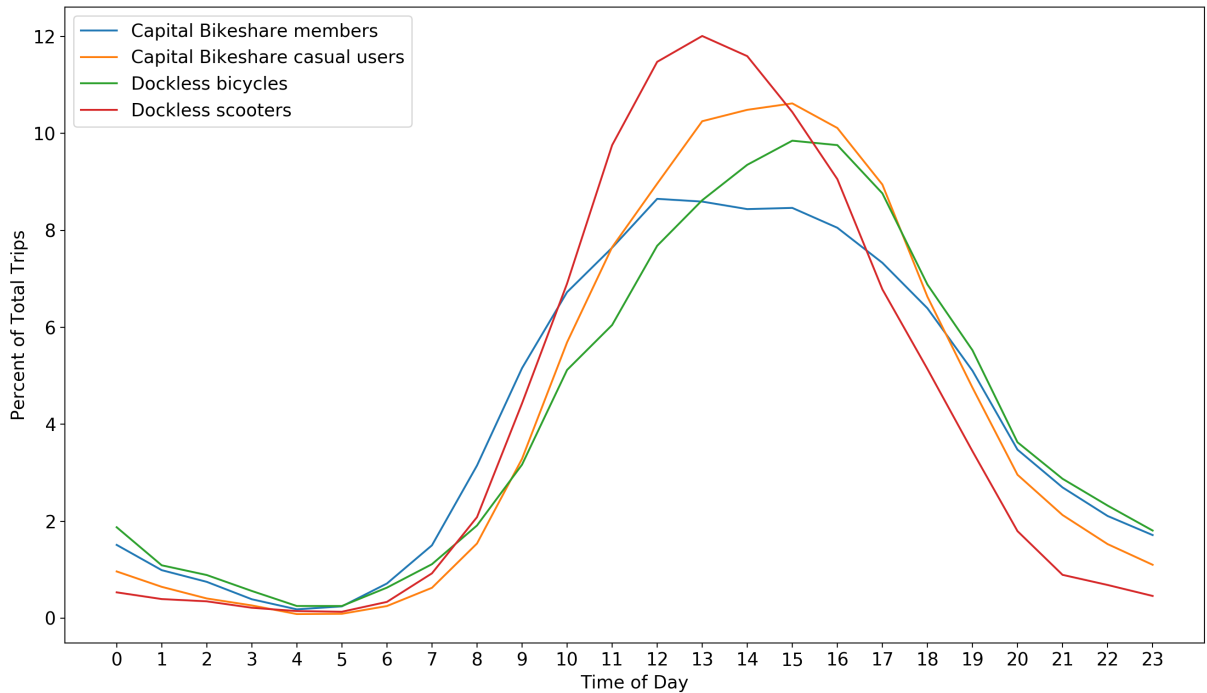

b) Weekends

**FIGURE 2 Temporal distributions of trips on weekdays and weekends for each mode**

The peak hours for dockless bicycles trips were at 8-10 a.m. and 5-7 p.m., lasting one hour longer than typical peak hours. This was also found in a research report published by Virginia Tech for a project done for the District Department of Transportation *(VirginiaTech, 2018)*. The afternoon peak was from 12-1 p.m., similar to the CaBi registered members. Morning peak for scooters was the same as dockless bicycles, 8-9 a.m., while afternoon peak for scooters started at 12 p.m. and lasted until 5 p.m. The highest peak of scooter trips was at 3 p.m. Since scooters require overnight charging, their rides were concentrated during daylight hours when they were readily available.

Weekend curves are somehow similar for all riders. Nearly 80% of the weekend rides for all operators were made in a period between 10 a.m. and 6 p.m. Ridership peaks differed slightly among different types of users we compared. For weekend trips, the peak for CaBi registered members was at 12 p.m., for scooters at 1 p.m., and for casual CaBi riders and dockless bicycle users at 3 p.m.

4.3. Trip duration

The duration of trips was similar for dockless bicycles and scooters, and significantly different for casual riders of CaBi. Figure 3 shows the band chart of trip duration for all the modes, with the lower line for the 25$^{th}$ percentile, middle line for 50$^{th}$ percentile, and upper for the 75$^{th}$ percentile. As it is indicated in Figure 3, CaBi members kept their trips short, with 75% of trips less than 15 minutes during most of the hours. Members made their shortest trips during morning hours between 5 a.m. to 9 a.m. and after 8 p.m., a trend that is also seen for other types of users. Similar to CaBi members, dockless bicycle and scooter riders had median trip durations of around 10 minutes. While dockless bikeshare and scooter users had a slightly longer trip duration in the afternoon, casual riders made the longest trips among all. In the middle of the day, the median trip duration for casual users was between 20 and 27 minutes, and only 25% of their trips were shorter than 15 minutes.

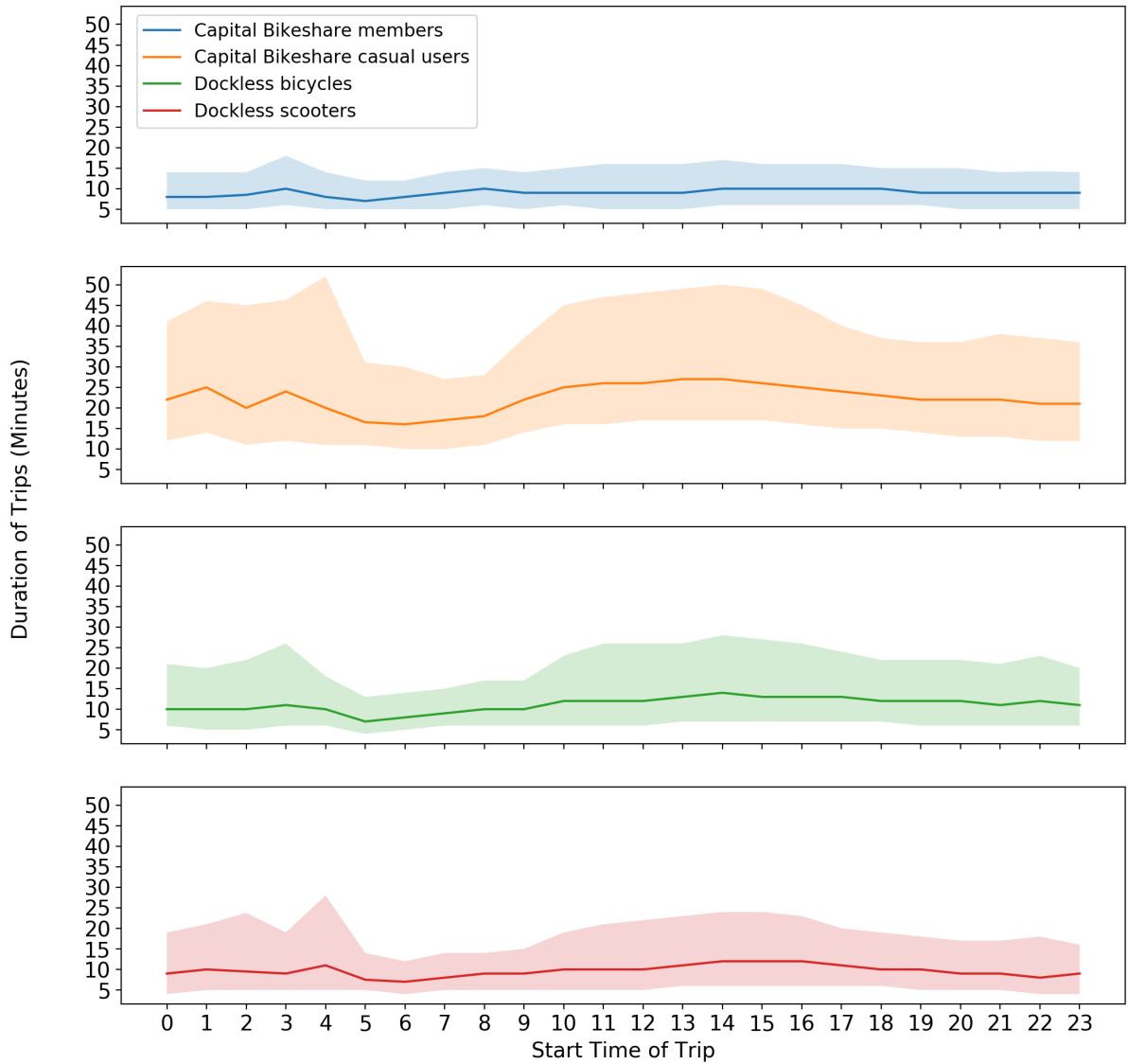

**FIGURE 3 Band chart for 25ᵗʰ percentile, 50ᵗʰ percentile, and 75ᵗʰ percentile of trip duration versus start time of trip for each mode**

### 4.4. Spatial distribution of the trips

Capital Bikeshare riders can start and end their trips only at stations, as opposed to the dockless system with the possibility of flexible start and end locations. Therefore, an aggregated level of analysis was chosen to make the comparative analysis of trips between the two systems feasible. Because of this limitation, a detailed density analysis, which would be plausible for dockless trips, was not conducted in this study. The 2013 single-member districts level (SMD) was used as the geographic unit of analysis and the trips were aggregated to SMDs based on their start/end locations. SMDs are similar in terms of population with each SMD residing about 2,000 people, but could have different sizes in terms of geographic area.

The spatial distribution of trips differed for the observed types of shared vehicles. To make the difference obvious, the start location of trips during morning peak (7-9 a.m.) was chosen as the most representative time window. Figure 4 shows the percent of trips started in the morning peak by users of different services. In total, CaBi stations were located in 146 out of 296 SMDs, and dockless bicycles and scooters riders were able to start from any SMD. In the mornings, CaBi members started from 137 SMDs and casual riders from 133 SMDs. Dockless bikeshare and scootershare trips started from 231 and 221 SMDs respectively. The dockless bicycle trips covered a larger territory than scooter trips; this could be partly because of the different fleet size between scooters and dockless bikes as well as different relocation/rebalancing strategies. The Southeast part of the District is underrepresented in the morning trips maps for all compared vehicles.

In the morning peak, casual CaBi riders and dockless bicycle users started and ended their trips mostly in downtown, while CaBi members and electric scooter riders mostly started from mixed-use neighborhoods and ended in downtown in the morning. Among all user types, CaBi's registered members showed the closest behavior to daily commuting pattern, with starting morning trips from the mixed-use and residential neighborhoods 1-3 miles away from downtown and making trips in the opposite direction during the afternoon peak. Casual CaBi riders, in contrast, were mostly concentrated around the National Mall and White House area. Although the percentage of casual trips per SMD does not exceed 2.5 in all but five SMDs, the percentage of casual trips started around National Mall dramatically increases: 10.4% for the SMD that covers governmental district Federal Triangle and the eastern part of the National Mall and 7.9% for the SMD covering White House and the rest of the Mall. Three percent of morning casual trips were started from Union Station, which is the second-busiest railroad station and headquarter of Amtrak, National Railroad Passenger Corporation (See Figure 4).

Dockless bikes' usage in the morning was also concentrated in the downtown area. The highest percentage of dockless bicycle trips, 4.2%, was in the SMD covering the National Mall, followed by SMDs in L'Enfant Plaza and Farragut neighborhoods where many jobs are concentrated. The concentration of trips around metro stations during the morning and afternoon peak hours supports the assumption of dockless system usage as a first-mile/last-mile solution complementing their Metro ride. However, a more detailed transit and bikeshare ridership data and additional research is required to confirm this assumption.

The scooter riders mostly started their morning trips in the mixed-use neighborhood to the north, east, and west of downtown. The largest percentage of trips started by scooters in the morning peak was 5.5% from U street NW, a mixed-use neighborhood three miles north from downtown, and from surrounding neighborhoods. In general, the trip distribution of scooters looked similar to CaBi members.

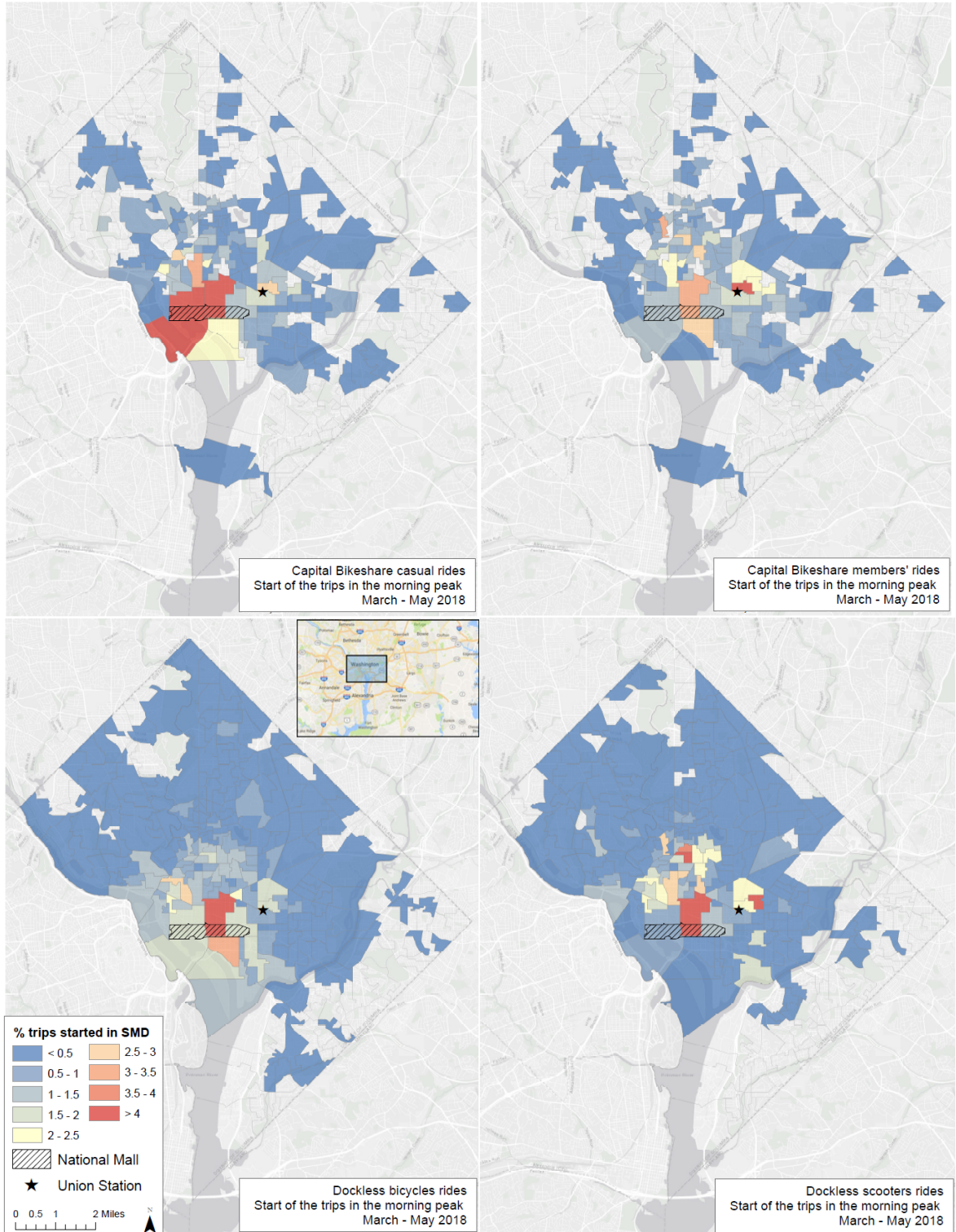

**FIGURE 4 Trip start locations in the morning peak for each mode (7-9 a.m.)**

## 5. Modeling approach

Two classification models were trained based on Capital Bikeshare's data to investigate the users' behavior of dockless system in Washington, D.C. A logistic regression model proposed by David Cox (1958) and a random forest model proposed by Breiman (2001). A random forest classifier consists of a collection of tree-structured classifiers which are created by randomly selecting feature vectors for randomly selected training datasets. Both of these models are widely used in the literature for classification purposes. There is no general rule that one performs better than the other. However, some studies suggested that the random forest will usually perform better on datasets with a higher ratio of features size to their training size (Couronné et al., 2018). Although for our dataset random forest showed a slightly better performance in terms of evaluation metric, we have included the result of both models. As the logistic regression model has the benefit of understanding the direction of association between the features and the predictor.

### 5.1. Evaluation

Precision, recall, and $F_1$ score are typically used as a measurement of accuracy for binary classifications.

$$Precision = tp/(tp + fp) \qquad (1)$$

$$Recall = tp/(tp + fn) \qquad (2)$$

$$F_1 = 2 * (Precision * Recall)/(Precision + Recall) \quad (3)$$

where:

      tp = true positive
      fp = false positive
      fn = false negative

Considering these measurements in the model selection process is especially important when the dataset to classify is not balanced, i.e., one of the classes is more prevalent than the others.

### 5.2. Preprocessing, feature selection and model selection

To train a classification model that has the capability of differentiating the trips made by members and casual riders of Capital Bikeshare, we first identify the trip characteristics that distinguish the types of users based on trips attributes that were relevant and available in both CaBi and dockless dataset. As shown by the comparative analysis, members and casual riders have different ridership behavior in terms of start/end time and location, day of the week, and trip duration. As a result, training for both logistic regression and random forest models was based on the following variables: day of week, start time, trip duration, start location, and end location. Several preprocessing steps are executed before training the model. Trip start time is converted to 30-minute intervals. Since model trained by CaBi data is used to classify the trips by dockless operators, we have used location of CaBi stations in District of Columbia for partitioning the region and identifying start/end locations. A Voronoi partition based on CaBi stations is used as the level of analysis for aggregating the data. This method is widely used in the literature for partitioning of plane into regions. McKenzie (2019) has used this method for comparing spatial distribution of dockless bikeshare and station-based bikeshare systems. As a result, the region is divided into 269 (number of CaBi stations in District of Columbia during this period) polygons. Figure 5 visualizes this partitioning. One hot encoding is done to convert all of the variables except trip duration into binary variables. Trip duration is used as a continuous variable that is rescaled into the range between 0 and 1 in both models.

Another preprocessing step used for this dataset is downsampling. The number of trips made by registered members of CaBi is significantly higher than the number of trips by casual users. As a result, downsampling is employed to avoid fitting the model in favor of the class that has the majority of trips. In this method, the size of the majority class is reduced by randomly selecting records until the size of two classes becomes equal. After downsampling is done, the final number of observations for each class (members and casual users) is 217,507. Random sampling was verified to have a similar distribution as the original dataset by checking the summary statics of the trips' attributes of the sample and the original dataset. The dataset was then divided into training and test sets. Here, 80% of the dataset was randomly chosen for training and the rest was kept for testing the model.

For the random forest model, several combinations of hyperparameters were tested to find the optimal hyperparamters. Both of the models were evaluated based on their $F_1$ score

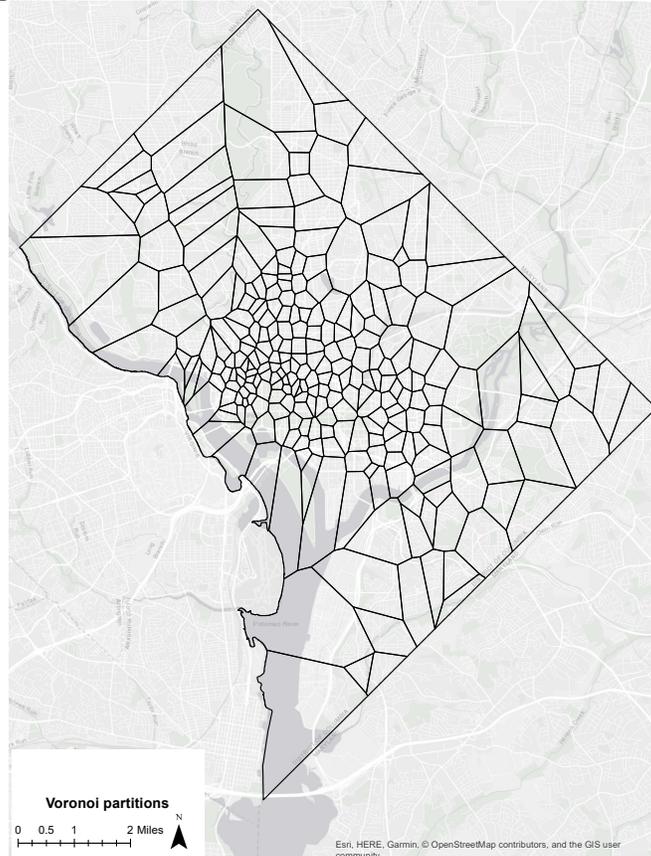

**FIGURE 5 Voronoi partition based on CaBi stations**

## 6. Modeling results

### 6.1. Modeling results for Capital Bikeshare

Table 1 shows the confusion matrix results of the trained model on the test dataset. Numbers in bold font represent the instances correctly classified by the trained model. For example, the trained random forest model correctly labeled 36,293 trips out of 43,374 trips by casual users. These numbers are slightly lower for the logistic regression.

**TABLE 1 Confusion Matrix Results for Logistic Regression and Random Forest (Test Data Set)**

| | Logistic Regression | | Random Forest | |
|---|---|---|---|---|
| Predicted label<br>Actual label | Casual | Member | Casual | Member |
| Casual | **33,363** | 10,011 | **36,293** | 7,081 |
| Member | 6,443 | **37,186** | 6,662 | **36,967** |

Table 2 shows precision, recall, and $F_1$ score based on the test dataset for the CaBi. Random forest model has a slightly better $F_1$ score (0.84 versus 0.81). This implies that the random forest model has a slightly better prediction performance compared to the logistic regression model. Please refer to the evaluation section (section 5.1) for a discussion of these performance measurements.

**TABLE 2 Precision, Recall, and $F_1$-scores for Logistic Regression and Random Forest (test set)**

| | Test sample | Precision | Recall | $F_1$-Score |
|---|---|---|---|---|
| *Logistic regression* | | | | |
| Casual | 43,374 | 0.84 | 0.77 | 0.80 |
| Member | 43,629 | 0.79 | 0.85 | 0.82 |
| Average/total | 87,003 | 0.81 | 0.81 | 0.81 |
| *Random forest* | | | | |
| Casual | 43,374 | 0.84 | 0.84 | 0.84 |
| Member | 43,629 | 0.84 | 0.85 | 0.84 |
| Total/average | 87,003 | 0.84 | 0.84 | 0.84 |

In the logistic regression model, variables with large positive coefficients and small negative coefficients are the most significant variables influencing the classification process. Looking at the coefficients in the logistic regression model results, we found that coefficients of trip duration (negative) and start time between 4:30 a.m. and 9:30 a.m. (positive) have the largest absolute values in the model. Since in the trained model outcome of 1 means being a member trip, this indicates that high duration trips decrease the likelihood of being a member trip whereas having the start time of trip between 4:30 a.m. and 9:30 a.m. increases the likelihood of being a member trip. This is intuitive as most of the trips taken by casual riders have higher duration and most of the trips by members happen during commute hours. Figure 6-a shows start/end polygons that have coefficients with high absolute value in the model. The ten blue polygons show locations that have negative coefficients for both start and end meaning that trips that start and end at these locations are made mostly by casual riders and ten yellow polygons show locations that have positive coefficients for both start and end meaning that trips that start and end at these locations are made mostly by members.

For the random forest model, feature importance was used as the measurement for how well that variable can distinguish the two types of users. Feature importance was obtained by calculating the total decrease in node impurity weighted by the probability of reaching that node averaged over all the trees. Figure 6-b shows start/end polygons with high feature importance. The following variables have the highest importance in distinguishing the two classes in the random forest model: trip duration, whether a trip happened on Saturday, and whether a trip happened on Sunday.

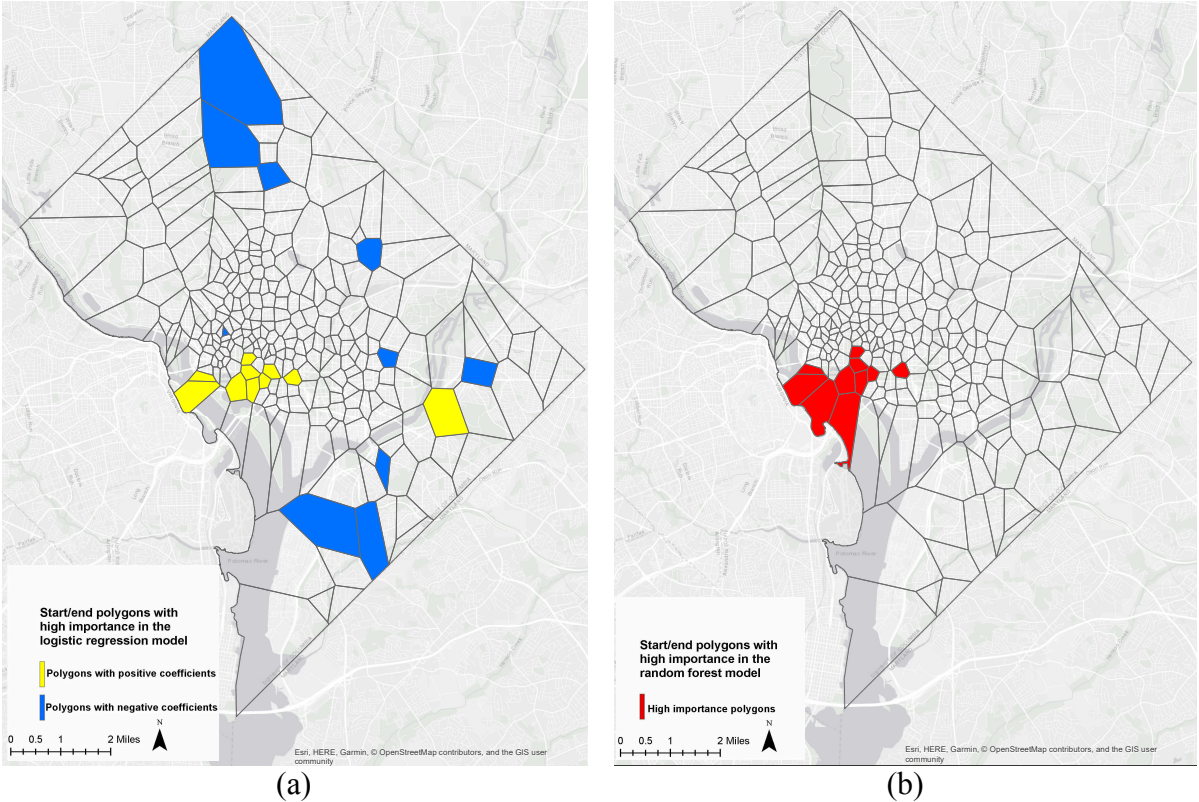

(a)                                                        (b)

**FIGURE 6 Start/end polygons with high importance a) Logistic regression model b) Random forest model**

## 6.2. Modeling results for dockless dataset

Since there is no ground truth for the dockless systems, we had to use the trained model on CaBi dataset for inferring the label of dockless trips. Both trained models were applied on dockless dataset to see what percentage of trips are going to be classified as casual trips and what percentage are going to be classified as member trips. Table 3 presents the number of trips classified as casual or member type and the percent of dominating type. As it indicates, 130,836 trips out of 187,909 total trips for dockless scooters were identified as members' trips by the random forest model. The logistic regression model labeled around 71.8% and 65.3% of dockless scooters' trips and dockless bikes' trips respectively as members' trips. For the random forest model, these numbers were 69.6% and 63.8% respectively. The results of both models show that most of the trips are classified as member trips. It is also shown in the features' importance of the trained models that duration of trips plays a significant role in distinguishing between CaBi members and casual riders. The summary statistics of the duration of trips for dockless bikes and scooters shows that dockless users behave more similarly to members than casual users. This is probably the reason a higher percentage of trips are labeled as members in our model.

**TABLE 3 Model Prediction for Dockless Bikes and Dockless Scooters**

|  | Logistic Regression | | Random Forest | | |
| --- | --- | --- | --- | --- | --- |
| Predicted behavior / Vehicle type | Casual | Member | Casual | Member | Total |
| Dockless bicycles | 24,774 | **46,816** (65.3%) | 25,894 | **45,696** (63.8%) | 71,590 |
| Dockless scooters | 52,822 | **135,087** (71.8%) | 57,073 | **130.836** (69.6%) | 187,909 |

## 7. Conclusions and study limitations

This study focuses on analyzing ridership data of three shared mobility modes. The results of this study could shed some light on the types of users these modes support and when and where they are used. The findings showed that in general, users of dockless bicycles and scooters tend to have shorter trips than casual CaBi users and slightly longer trips than CaBi registered members. Temporal distribution of trips showed that some characteristics of dockless trips are similar to CaBi members' trips (esp. during morning peak), and some of casual riders' trips (extended afternoon peak). Scooters riders showed commuters' behavior, starting their morning trips from mixed-use neighborhoods similar to Capital Bikeshare members. Dockless bicycles' trips were concentrated in downtown, similar to Capital Bikeshare casual riders, but also in employment areas. This suggests that the dockless system complements the conventional bikeshare system in Washington, D.C., even though there is a difference observed between the behavior of dockless bikes' users vs. scooters' users in comparison with conventional bikes users. Modeling results showed that ridership behavior of dockless bicycles and scooters is more closely aligned with Capital Bikeshare members patterns than with casual. According to the random forest model analysis, 63.8% of dockless bicycles' trips and 69.6% of scooters' trips were more similar to CaBi members' trips than to casual users.

Findings of this study could potentially be helpful for planners and policy makers currently assessing dockless bicycle and scooter implementation. Findings suggest the use of dockless systems demonstrates characteristics of both dock-based member and casual user behavior, and in turn, both commuter and recreational uses. However, by comparing the percentage of trips taken by registered members of CaBi (73.3%), we can conclude dockless systems are more used for recreational purposes. The distinctions in behavior may be useful to municipalities that have existing bikeshare systems or are considering launching dockless bikeshare or scootershare. Policymakers may also acknowledge the benefits of dockless systems when crafting rules and regulations by focusing on the equitable distribution of vehicles, the fidelity of locating systems, and controlling for their negative externalities thereby maximizing their benefit to the public. From the operator's perspective, this indicates offering long-term membership options for the users might be beneficial.

Despite very interesting results in the present study, it has many limitations. First, trips' start/end locations are influenced by the rebalancing pattern of each operator as well as the destination locations of previous users for the dockless systems' trips and by the bicycle availability in stations for the conventional system. It may not entirely show the desired start location of the users, although it shows the desired destination for the dockless systems. For electric scooters, the starting points may be skewed by night charging locations, especially for the morning trips. Second, the duration of trips made by electric scooters may be shorter because of their higher speed and insensitivity to the topography. Another limitation is a significantly different number of bicycles or scooters in the companies' fleets. While Capital Bikeshare operated with nearly 4,000 bicycles, each dockless bikeshare and scootershare companies could have up to 400 vehicles in Spring 2018. These limitations may slightly distort the comparative analysis and modeling results in relation to starting points for dockless bicycles and scooters and trips duration of scooters. Future research is needed to address these issues. Moreover, considering the limited period of dockless bikeshare and scootershare operation in the United States, further research may be useful to compare trips characteristics in different seasons and over an entire year.

**Acknowledgement**

This research is partially funded by National Center for Smart Growth at the University of Maryland, College Park. The authors would like to thank the District Department of Transportation (DDOT) for providing data and support for this study. We would also like to thank Sean Thomas Burnett for his help in editing and providing helpful insights into this research. Authors are solely responsible for all the statements in this paper.